\pdfoutput=1

\documentclass[11pt,twoside,a4paper,cmspaper,final,collab]{cms-tdr}
\def\svnVersion{ffabd32-D}\def\svnDate{2019/12/17}\def\cmsCernNoTag{CERN-EP-2019-196}\def\cmsCernDate{\today}\def\cmsMessage{"Published in Physical Review Letters as \href{http://dx.doi.org/10.1103/PhysRevLett.124.041803}{\doi{10.1103/PhysRevLett.124.041803}.}"}

\begin{document}\cmsNoteHeader{SUS-19-002}

\hyphenation{had-ron-i-za-tion}
\hyphenation{cal-or-i-me-ter}
\hyphenation{de-vices}
\newcommand\sqrts{\sqrt{s}}
\newcommand\deltm{\Delta m}
\newcommand\scriptr{\mathcal{R}}
\newcommand\jrm{\textrm{j}}
\newcommand{\cmp}{\HepParticle{\PSGc}{1}{\mp}\xspace}
\newcommand\scriptb{\mathcal{B}}
\newcommand\abseta{\abs{\eta}}
\newcommand\Deltphi{\Delta\phi}
\newcommand\jisr{\textrm{j}_{\text{ISR}}}
\newcommand\qcd{\text{QCD}}
\ifthenelse{\boolean{cms@external}}{\providecommand{\cmsLeft}{upper\xspace}}{\providecommand{\cmsLeft}{left\xspace}}
\ifthenelse{\boolean{cms@external}}{\providecommand{\cmsRight}{lower\xspace}}{\providecommand{\cmsRight}{right\xspace}}
\newlength\cmsFigWidth
\ifthenelse{\boolean{cms@external}}{\setlength\cmsFigWidth{\columnwidth}}{\setlength\cmsFigWidth{0.75\textwidth}}
\title{Search for supersymmetry with a compressed mass spectrum in events with a soft \texorpdfstring{\PGt}{tau} lepton, a highly energetic jet, and large missing transverse momentum in proton-proton collisions at \texorpdfstring{$\sqrts=13$\TeV}{sqrt(s) = 13 TeV}}

\author*[inst1]{Alfredo Gurrola}
\date{\today}

\abstract{
The first search for supersymmetry in events with an experimental signature of one soft, hadronically decaying \Pgt lepton, one energetic jet from initial-state radiation, and large transverse momentum imbalance is presented. These event signatures are consistent with direct or indirect production of scalar \Pgt leptons ($\PSgt$) in supersymmetric models that exhibit coannihilation between the \PSgt and the lightest neutralino (\PSGczDo), and that could generate the observed relic density of dark matter. The data correspond to an integrated luminosity of 77.2\fbinv of proton-proton collisions at $\sqrts=13\TeV$ collected with the CMS detector at the LHC in 2016 and 2017. The results are interpreted in a supersymmetric scenario with a small mass difference ($\deltm$) between the chargino (\PSGcpmDo) or next-to-lightest neutralino (\PSGczDt), and the \PSGczDo. The mass of the \PSgt is assumed to be the average of the \PSGcpmDo and \PSGczDo masses. The data are consistent with standard model background predictions. Upper limits at 95\% confidence level are set on the sum of the \PSGcpmDo, \PSGczDt, and \PSgt production cross sections for $\deltm(\PSGcpmDo, \PSGczDo) = 50\GeV$, resulting in a lower limit of 290\GeV on the mass of the \PSGcpmDo, which is the most stringent to date and surpasses the bounds from the LEP experiments.
}

\hypersetup{%
pdfauthor={CMS Collaboration},%
pdftitle={Search for supersymmetry with a compressed mass spectrum in events with a soft tau lepton, a highly energetic jet, and large missing transverse momentum in proton-proton collisions at sqrt(s) = 13 TeV},%
pdfsubject={CMS},%
pdfkeywords={CMS, physics, supersymmetry}}

\maketitle

Supersymmetry (SUSY)~\cite{PhysRevD.3.2415, JETP.13.1971, FERRARA1974413, WESS197439, PhysRevLett.49.970, BARBIERI1982343, PhysRevD.27.2359} is a theoretical extension of the standard model (SM)
that could describe the particle nature of dark matter (DM) and solve the gauge hierarchy problem. In SUSY models assuming $\scriptr$ parity~\cite{Dimopoulos:1995mi} conservation, if the
lightest neutralino (\PSGczDo) is the lightest
SUSY particle, it is neutral, stable, and could have undergone annihilation-production interactions with SM particles in the early universe to give the DM relic density observed
today~\cite{Spergel:2003cb,Ade:2015xua}. In models with a bino (\PZ-like) \PSGczDo, these interactions alone are insufficient to produce the correct DM relic abundance. As such, a model of coannihilation (CA) can be introduced, where CA refers to the interaction of \PSGczDo with another SUSY particle resulting in the production of SM particles~\cite{PhysRevD.43.3191}.

This Letter describes a search for the production of stau particles (\PSgt), SUSY partners of the \Pgt lepton, considering a mass difference ($\deltm$) between the \PSGczDo and \PSgt of $\leq{50}\GeV$. These scenarios are motivated by models including \PSgt-\PSGczDo CA~\cite{Arnowitt:2006bb,Khotilovich:2005gb, Carena:2012gp, Aboubrahim:2017aen, Duan:2018cgb, PhysRevLett.100.231802, VBFPheno, VBFPheno2}, where the calculated relic DM density is consistent with that measured by the WMAP and Planck Collaborations~\cite{Spergel:2003cb,Ade:2015xua}. The CA cross section is exponentially enhanced by small $\deltm(\PSgt, \PSGczDo)$.

In proton-proton ($\Pp\Pp$) collisions at the LHC, \PSgt particles can be produced directly in pairs or in decays of heavier SUSY particles. The \PSgt can decay to a \Pgt lepton and
\PSGczDo. The analysis described in this Letter requires an extra jet ($\jrm$) from initial-state radiation (ISR). The recoil effect from the ISR jet facilitates the detection of momentum
imbalance and identification of the low-energy (soft) \Pgt lepton decay products~\cite{ISRStauPheno,VBFPheno,VBFPheno2,VBFSbottom,VBFStop,VBFSlepton,VBFZprime,VBFHN,VBFY2}. Thus, this
analysis focuses on $\Pp\Pp \to \PSgt\PSgt\jrm$ production and indirect \PSgt
production via decays of the lightest chargino (\PSGcpmDo) or the next-to-lightest neutralino (\PSGczDt) in processes like $\Pp\Pp \to \PSGcpmDo\cmp\jrm \to \PSgt\PSgt\Pgngt\Pgngt\jrm \to
\Pgt\PSGczDo\Pgt\PSGczDo\Pgngt\Pgngt\jrm$ and $\Pp\Pp \to \PSGcpmDo\PSGczDt\jrm \to \PSgt\Pgngt\PSgt\Pgt\jrm \to \Pgt\PSGczDo\Pgngt\Pgt\PSGczDo\Pgt\jrm$, which can be the dominant production
mechanisms for \PSgt via decays of heavier SUSY particles.
While these processes yield final states with multiple \Pgt leptons, the average transverse momentum (\pt) of the \Pgt leptons is
$\deltm$/2 and below the reconstruction threshold in the $\deltm(\PSgt, \PSGczDo) \leq 50\GeV$ scenarios.
The visible decay products of the \Pgt leptons have lower \pt than the decaying
particles, so it is difficult to identify more than one \Pgt lepton in a signal event. Furthermore, leptonic decays of \Pgt leptons have a smaller branching fraction ($\scriptb$) than
hadronic decays (\tauh), and, on average, smaller visible \pt. Electrons and muons from such decays are also indistinguishable from prompt production of electrons and muons. Hence, we
search for events with exactly one soft \tauh candidate and missing transverse momentum recoiling against a high-\pt ISR jet.

The strategy above allows this analysis to probe the \PSgt-\PSGczDo CA region with $\deltm(\PSgt, \PSGczDo) \leq 50\GeV$. This is the first collider search for compressed SUSY spectra using
this strategy. Earlier searches from the CMS and ATLAS Collaborations~\cite{CMSEWK, CMSsoftdilepton, Aad:2014nua, ATLASTsoftdilepton, Khachatryan:2015kxa, SUS17007, SUS18006} that relate to
the scenarios in this Letter produced weaker results than those from the LEP experiments~\cite{Heister:2001nk, Abdallah:2003xe, Achard:2003ge, Abbiendi:2003ji}. Data
collected in 2016 and 2017 with the CMS experiment~\cite{CMSdetector} in $\Pp\Pp$ collisions at $\sqrts = 13\TeV$ is used. The data sample corresponds to an integrated luminosity of 77.2\fbinv.

The central feature of the CMS apparatus~\cite{CMSdetector} is a superconducting solenoid of 6\unit{m} internal diameter, providing a magnetic field of 3.8\unit{T}. Within the solenoid volume are a 
silicon pixel and strip tracker, a lead tungstate crystal electromagnetic calorimeter (ECAL), and a brass and scintillator hadron calorimeter, each composed of a barrel and two endcap sections. 
Forward calorimeters extend the pseudorapidity ($\eta$) coverage of the barrel and endcap detectors up to $\abseta<5.2$. Muons are measured in gas-ionization detectors embedded in the steel 
flux-return yoke outside the solenoid. A detailed description of the CMS detector can be found in Ref.~\cite{CMSdetector}.

Events are reconstructed from particle candidates (electrons, muons, photons, and hadrons) identified using the particle-flow (PF) algorithm~\cite{CMS-PRF-14-001}. The algorithm combines information from all detectors to classify final-state particles produced in the collision. Jets are clustered using the anti-\kt clustering algorithm~\cite{Cacciari:2008gp, Cacciari:2011ma} with a distance parameter of 0.4. Identification criteria are applied to jet candidates to remove anomalous effects from the calorimeters~\cite{Sirunyan:2019kia}. For jets with $\pt>30\GeV$ and $\abseta<2.4$, the identification efficiency is $>$99\%~\cite{CMS-PAS-JME-14-003}.

The jet energy scale and resolution are corrected depending on the \pt and $\eta$ of the jet~\cite{CMS:JetResol}. Jets originating from the hadronization of \PQb quarks are identified using the combined secondary vertex algorithm \cite{CMSbjet2}. This analysis uses the loose working point of the algorithm, which has an identification efficiency of 80\% for \PQb jets and a light-flavor quark or gluon misidentification rate of 10\%.

Electrons and muons are used in control samples and as vetoes in the signal sample selection.  Electrons are reconstructed and identified combining information from the ECAL and the tracking 
system~\cite{Khachatryan:2015hwa}. Muons are reconstructed using the tracker and muon chambers, and requiring consistency with 
low-energy measurements in the calorimeters~\cite{Sirunyan:2018fpa}. For this analysis, the electron (muon) identification efficiency is 85 (96)\%, for leptons with $\pt >10\GeV$ and $\abseta<2.1$.

Hadronic decays of \Pgt leptons are reconstructed and identified using the hadrons-plus-strips algorithm \cite{Sirunyan:2018pgf}, designed to optimize \tauh reconstruction by considering specific
\tauh
decay modes. To suppress backgrounds from light-flavor quark or gluon jets, \tauh candidates are required to pass a threshold value of a multivariate discriminator that takes variables related to
isolation and \Pgt lepton lifetime as input.
The tight isolation working point is used, which results in a \tauh identification efficiency of 55\% for this analysis, and a 0.2--5\% probability for a jet to be misidentified as a \tauh, depending on the \pt and $\eta$ values of the \tauh candidate~\cite{Sirunyan:2018pgf}. The \tauh candidates are subject to additional requirements, based on consistency among measurements in the tracker, calorimeters, and muon detectors, to distinguish them from electrons and muons.

The missing transverse momentum \ptvecmiss is the negative vector \pt sum of all PF candidates.  Its magnitude is \ptmiss. Production of undetected particles such as SM neutrinos and the 
\PSGczDo is inferred from the measured \ptmiss~\cite{SUS14019,VBFADM}. The jet corrections described are propagated as corrections to \ptmiss, which improves agreement in \ptmiss between
simulation and data.

The dominant SM background processes contributing to the search are $\PW$/$\PZ$ boson production in association with jets ($\PW$+jets and $\PZ$+jets), top quark pairs (\ttbar), and quantum 
chromodynamics (QCD) multijet processes. The contributions of $\PW$+jets and $\PZ$+jets events contain genuine \tauh candidates, energetic jets, and \ptmiss from neutrinos. Background from \ttbar 
events is characterized by two \PQb quark jets in addition to a genuine \tauh. QCD multijet events are characterized by jets misidentified as \tauh, and the estimated yield of this background 
is derived from data.

Simulated samples for $\PZ$+jets, $\PW$+jets, $\ttbar$+jets, and single top quark events are produced with the \MGvATNLO 2.6.0 program~\cite{aMCatNLO} at leading order (LO) precision. The
LO \PYTHIA generator is used to model diboson (VV) processes. Two sets of signal event samples are generated using \MGvATNLO 2.3.3 at LO precision.
The first set considers the sum of $\PSGcpmDo\cmp$, $\PSGcpmDo\PSGczDt$, and $\PSgt\PSgt$ production with up to two jets. The $\PSgt\PSgt$ process represents $<$2\% of the total cross section.
Models with a bino \PSGczDo and wino (\PW-like) \PSGczDt and \PSGcpmDo are
considered. We assume a simplified model scenario~\cite{Alves:2011wf} with a left-handed \PSgt, $\scriptb(\PSGczDt\to{\Pgt\PSgt}\to{\Pgt\Pgt\PSGczDo})
=\scriptb(\PSGcpmDo\to{\Pgngt\PSgt}\to{\Pgngt\Pgt\PSGczDo}) = 100\%$, and $m(\PSGcpmDo) = m(\PSGczDt)$. This set of samples is motivated by the importance of the chargino-neutralino sector in
connecting SUSY models and DM. We refer to this model as SUSY signal model 1 (SSM1). The second set considers direct production of left-handed \PSgt pairs with up to
two jets. Although the search for direct \PSgt production with $\deltm(\PSgt,\PSGczDo) \leq 50\GeV$ is challenging because of the small production cross section and low signal acceptance,
this set of samples is included to highlight the improved sensitivity in this analysis, compared to previous non-ISR
searches~\cite{Heister:2001nk,Abdallah:2003xe,Achard:2003ge,Abbiendi:2003ji,Aad:2014,PhysRevD.93.052002,Khachatryan2017,Khachatryan:2015kxa,Sirunyan2018,ATLASTDirectStau}. This second set of 
samples allows for reinterpretation in other scenarios
with \PSgt-like particles. We refer to this model as SUSY signal model 2 (SSM2). It is noted that the masses of $\PSGcpmDo/\PSGczDt$ are sufficiently large (10\TeV) to be considered decoupled
in SSM2.  The \MGvATNLO generator is interfaced with \PYTHIA 8.212~\cite{Sjostrand:2014zea} using the CUETP8M1 and CP5 tunes~\cite{Khachatryan:2015pea,Sirunyan:2019dfx} for parton shower and
fragmentation in the 2016 and 2017 simulated samples, respectively. The NNPDF3.0 LO and NLO~\cite{Ball:2014uwa} parton
distribution functions (PDFs) are used in the event
generation. The CMS detector response is simulated using the \GEANTfour~\cite{Agostinelli:2002hh} package for background samples, and the CMS fast simulation
package~\cite{Abdullin:2011zz} for signal samples. To model the effect of additional $\Pp\Pp$ interactions within the same bunch crossing or nearby bunch crossings, minimum bias events
generated with \PYTHIA are added to the simulated samples with a frequency distribution per bunch crossing weighted to match that observed in data. MC background yields are normalized to the
integrated luminosity using next-to-next-to-leading order (NNLO) or next-to-leading order (NLO) cross sections, while signal production cross sections are calculated at NLO with next-to-leading
logarithmic (NLL) soft-gluon resummation calculations~\cite{Debove:2010kf,Fuks:2012qx,Fuks:2013vua,Fiaschi:2018hgm}.

Events are recorded using a \ptmiss trigger~\cite{Khachatryan:2016bia}. The trigger efficiency is measured using data events with one muon, resulting in a sample enriched in $\PW$+jets events (95\% 
purity in simulation). 
Selected events are required to have $\ptmiss > 230\GeV$, where the trigger is fully efficient, and exactly one identified \tauh candidate with $\abseta<2.1$ and $20 < \pt(\tauh) < 40\GeV$. The requirement of exactly one \tauh candidate and the upper limit on \pt reduce the $\PW$+jets, $\PZ$+jets, and $\ttbar$+jets backgrounds. The highest-\pt jet is referred to as the ISR jet ($\jisr$) and is required to satisfy $\pt > 100\GeV$ and $\abseta<2.4$. The absolute difference in the azimuthal angle ($\phi$) between the ISR jet and \ptvecmiss is required to be greater than 0.7 radians ($\abs{\Deltphi(\jisr,\ptvecmiss)} > 0.7$ radians) to reduce QCD multijet events containing large \ptmiss from jet mismeasurements. To reduce background processes with top quarks, events with \PQb-tagged jets are rejected. Events with well-identified and isolated electrons or muons with $\pt > 10\GeV$ and $\abseta<2.1$ are rejected.

The transverse mass of the selected \tauh candidate and the \ptvecmiss, defined as
\begin{equation*}
\mT(\ptvecmiss, \tauh)= \sqrt{{2 \ptmiss \pt(\tauh) (1-\cos\Deltphi(\ptvecmiss, \tauh))}},
\end{equation*}
is the main observable to search for the presence of signal events. The \mT in signal events probes the SUSY mass scale, and is expected to be larger on average than for the backgrounds. The strategy is to search for a broad enhancement in the high-\mT part of the spectrum.

The yield and \mT shape of the QCD multijet background are estimated from data using control regions (CRs) enriched in QCD multijet events and with negligible signal contamination. MC simulations 
are used to extrapolate the $\PW/\PZ$+jets and $\ttbar$+jets background yields from a CR to the signal region (SR) and to model \mT shapes. The agreement between data and simulation in these CRs is 
used to validate the modeling of the \tauh selections and to measure data-to-simulation scale factors to correct the modeling of the ISR jet and the \ptmiss. To calculate the correction factor, 
contributions from nontargeted backgrounds are subtracted from data. The uncertainty in these background processes is propagated to the final systematic uncertainty in the background predictions. 
Small contributions from single top quark and diboson production are estimated using simulation.

The correct modeling in the simulation of background events, in particular the $\PW/\PZ$+jets processes, can be affected by requiring an ISR jet. This modeling is studied using a $\PZ(\to
\PGm \PGm)$+jets CR in data. This CR provides a measurement of the \pt spectrum resulting from a high-\pt ISR jet, decoupling the effects of ISR modeling from the measurement of \ptmiss.
The \pt of the \PZ boson is measured by vectorially summing the transverse momenta of the two muons from the \PZ decay. The ratio between data and simulation in the $\pt(\PGm\PGm)$ distribution is
used to obtain \pt-dependent correction factors, ranging from 0.79 to 1.12.  The factors are validated using a $\PW(\to\PGm\PGnGm)$+jets enriched sample. After applying these
correction factors, we find agreement between the observed and predicted yields and shapes of distributions. These ISR correction factors are applied to all Drell--Yan processes, including the
$\PW/\PZ$+jets backgrounds and signal processes.

A $\PZ(\to \tauh \tauh)$+jets CR is defined to study the modeling of \tauh reconstruction and identification. The CR is obtained by requiring two \tauh candidates with $\pt > 60\GeV$ and
$\abseta<2.1$, selected by a dedicated $\tauh\tauh$ trigger~\cite{EXO16016,EXO17016,ZpToTauTau,Khachatryan:2015kxa}. The two \tauh candidates of a pair must have opposite electric charge and a
reconstructed mass
between 50 and 100\GeV, and all other
requirements are the same as for SR events. The contribution of QCD multijet events in the $\PZ(\to \tauh \tauh)$+jets CR is estimated from data using CRs obtained with \tauh pairs with the same
electric charge.  The transfer factor between same- and opposite-sign events is calculated using events with loosened \tauh isolation requirements and $m(\tauh \tauh) >100\GeV$. Correction factors
of $0.92 \pm 0.05$ and $0.95 \pm 0.04$ for $\PZ(\to \tauh \tauh)$+jets are measured in this CR for the 2016 and 2017 data sets, respectively. The uncertainties are purely statistical. These
correction factors
are used to scale the $\PZ(\to \tau \tau)$+jets prediction in the SR.

The contribution from \ttbar events in the SR is less than 15\% of the total expected background. Correction factors of $0.94 \pm 0.05$ and $0.95 \pm 0.04$ are measured for the 2016 and 2017 data
sets, respectively, in a
CR obtained by selecting events with two \PQb-tagged jets and one \tauh candidate with tighter isolation requirements with respect to the SR. These requirements allow for a \ttbar CR sample with high purity. The correction factor is applied to scale the prediction of \ttbar events in the SR.

A CR enriched with QCD multijet events (CR$_{\qcd}$) is obtained by requiring the same criteria for the SR but selecting \tauh candidates that fail the tight and pass the loose \tauh isolation. The contribution from nonmultijet events is subtracted using simulation, adjusted for the scale factors discussed above. The shape and normalization of the multijet background in the SR are predicted by multiplying the data yields in CR$_{\qcd}$ with transfer factors (``tight-to-loose'' ratios) to account for the isolation efficiency. The $\pt(\tauh)$-dependent transfer factors are derived in a $\PW(\to \PGm\PGnGm)+$\tauh CR, where the \tauh is a misidentified jet. These transfer factors, which range from 0.2 to 0.4, are validated in a region enriched in QCD multijet events by inverting the $\Deltphi(\jisr,\ptvecmiss)$ requirement.

A major source of systematic uncertainty is the closure of the background estimation methods, where closure refers to tests (on data and simulation) which demonstrate that the background
determination techniques reproduce the expected background distributions in both rate and shape within the statistical uncertainties. The background estimation uncertainty from the closure tests is
2--6\% for nonmultijet
backgrounds. For the QCD multijet background, this uncertainty is determined by the deviation of the tight-to-loose ratios obtained in a $\PZ(\to
\PGm\PGm)+$\tauh CR, where the
\tauh is a misidentified jet, from those in the $\PW(\to \PGm\PGnGm)+$\tauh region. This uncertainty depends on \pt(\tauh) and varies from 4 to 29\%. Shape-based systematic uncertainties from the use of ISR correction factors are determined by varying these factors by $\pm$1 standard deviation of their uncertainty and examining effects on the \mT distribution. This uncertainty is a few percent at low \mT and 15\% at high \mT. Although the corrected background \mT shapes are consistent with the data distributions within statistical uncertainties, data-to-simulation ratios of the \mT distributions are fit with a first-order polynomial, and the deviation of the fit from unity, as a function of \mT, is taken as an uncertainty in the shape. This results in up to 20\% uncertainty in a given \mT bin.

The signal and background yields estimated from simulation are affected by similar sources of systematic uncertainty, with small differences between the 2016 and 2017 data sets.  The uncertainty
from the \tauh
identification and isolation requirements ranges between 6 and 9\%, depending on the year and process~\cite{Sirunyan:2018pgf}. Efficiencies for the electron and muon reconstruction,
identification, and
isolation requirements are considered because of the extra lepton vetoes in the SR and their use in the CRs~\cite{Khachatryan:2015hwa,Sirunyan:2018fpa, CMS:2011aa}, with an uncertainty of $\leq$1\%. The \ptmiss scale uncertainties due to the jet energy scale (2--5\% depending on $\eta$ and \pt) result in an uncertainty of 1--3\% depending on \mT. The event acceptance for the ISR selection depends on the reconstruction and identification efficiencies and the energy scale of jets. A $\ptmiss$-dependent uncertainty in the measured trigger efficiency results in a 3\% uncertainty. The uncertainty in event acceptance from the PDF set used in simulation is evaluated in accordance with the PDF4LHC recommendations~\cite{PDF4LHCRunII} by comparing results using the CTEQ6.6L, MSTW08, and NNPDF10  PDF sets~\cite{Nadolsky:2008zw,mrst2006,nnpdf10} with those from the default PDF set. A systematic uncertainty in the signal accounts for differences between the fast and \GEANTfour simulations, which depends on \mT and varies from 3 to 11\%. The uncertainty in the integrated luminosity corresponds to 2.5~\cite{CMS-PAS-LUM-17-001} and 2.3\%~\cite{CMS-PAS-LUM-17-004} for the 2016 and 2017 data, respectively.

\begin{figure}[h!]\centering
  \includegraphics[width=\cmsFigWidth]{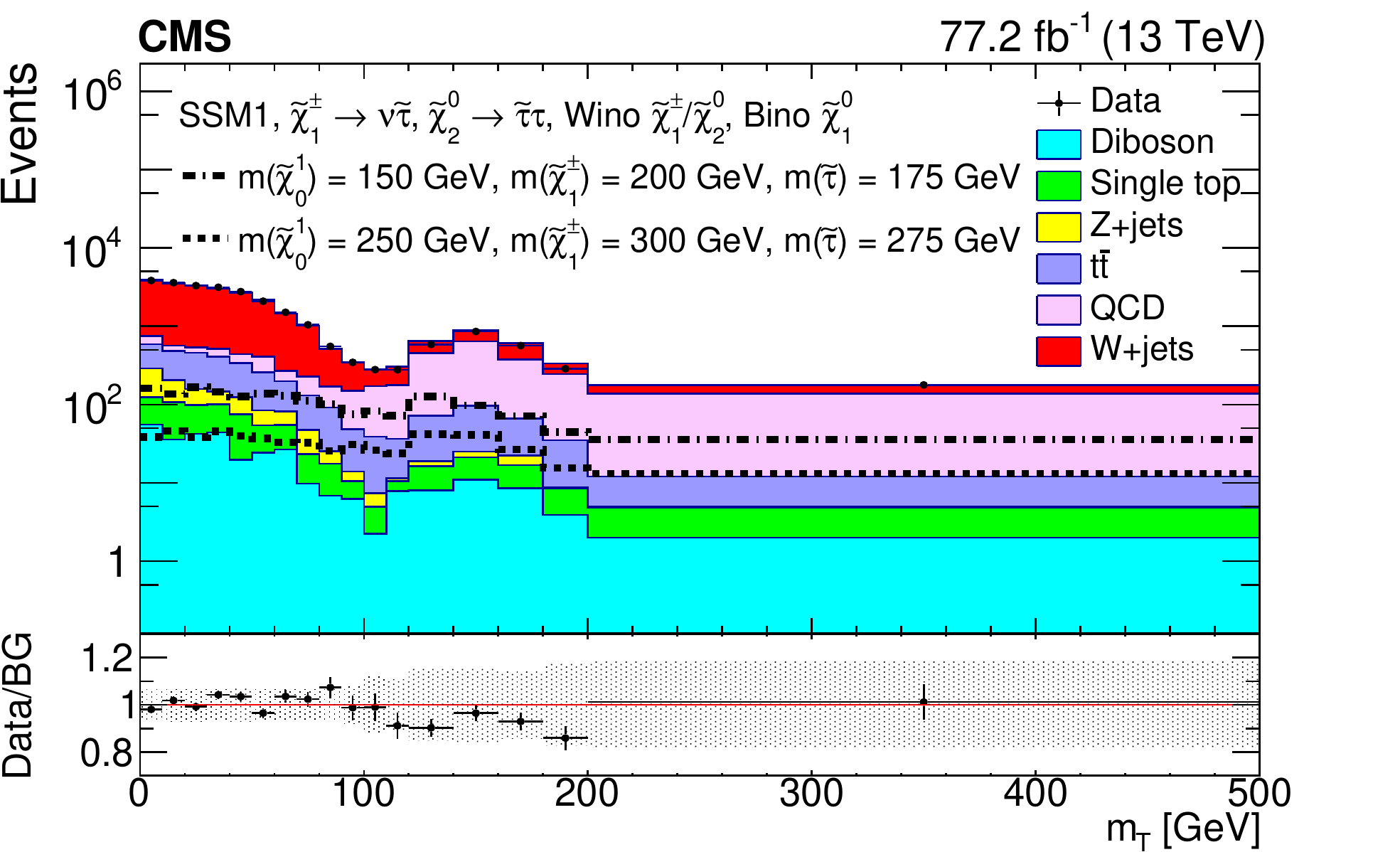}
 \caption{The \mT distribution for SR events with 2016 plus 2017 data. In the upper panel, the solid colors correspond to the expected
background processes, the black dots to the observed data, and the dashed lines to the expected signal from simulation. The lower panel shows the ratio between the observed data and
the total expected pre-fit
background (BG). The shaded band corresponds to the total pre-fit uncertainty on the BG prediction, while the error bars on the black dots represent the statistical uncertainties on the data
yields.}
 \label{fig:mTSR}
\end{figure}

Figure~\ref{fig:mTSR} shows the $\mT(\ptmiss, \tauh)$ distribution for events in the SR. The binning used in Fig.~\ref{fig:mTSR} is optimized to achieve the best discovery potential for the SSM1 
scenarios, resulting in bins of 10\GeV width between \mT of 0 and 120\GeV, bins of 20\GeV width between \mT of 120 and 200\GeV, and one bin of 300\GeV width for $\mT > 200\GeV$. For a SSM1 benchmark sample with $m(\PSGcpmDo) = 200\GeV$, $m(\PSgt) = 175\GeV$, and $m(\PSGczDo) = 150\GeV$, the signal-to-background ratio ranges from $\approx$1/25 at low \mT to $\approx$1/3 at high \mT.

\begin{figure}\centering
  \includegraphics[width=0.48\textwidth]{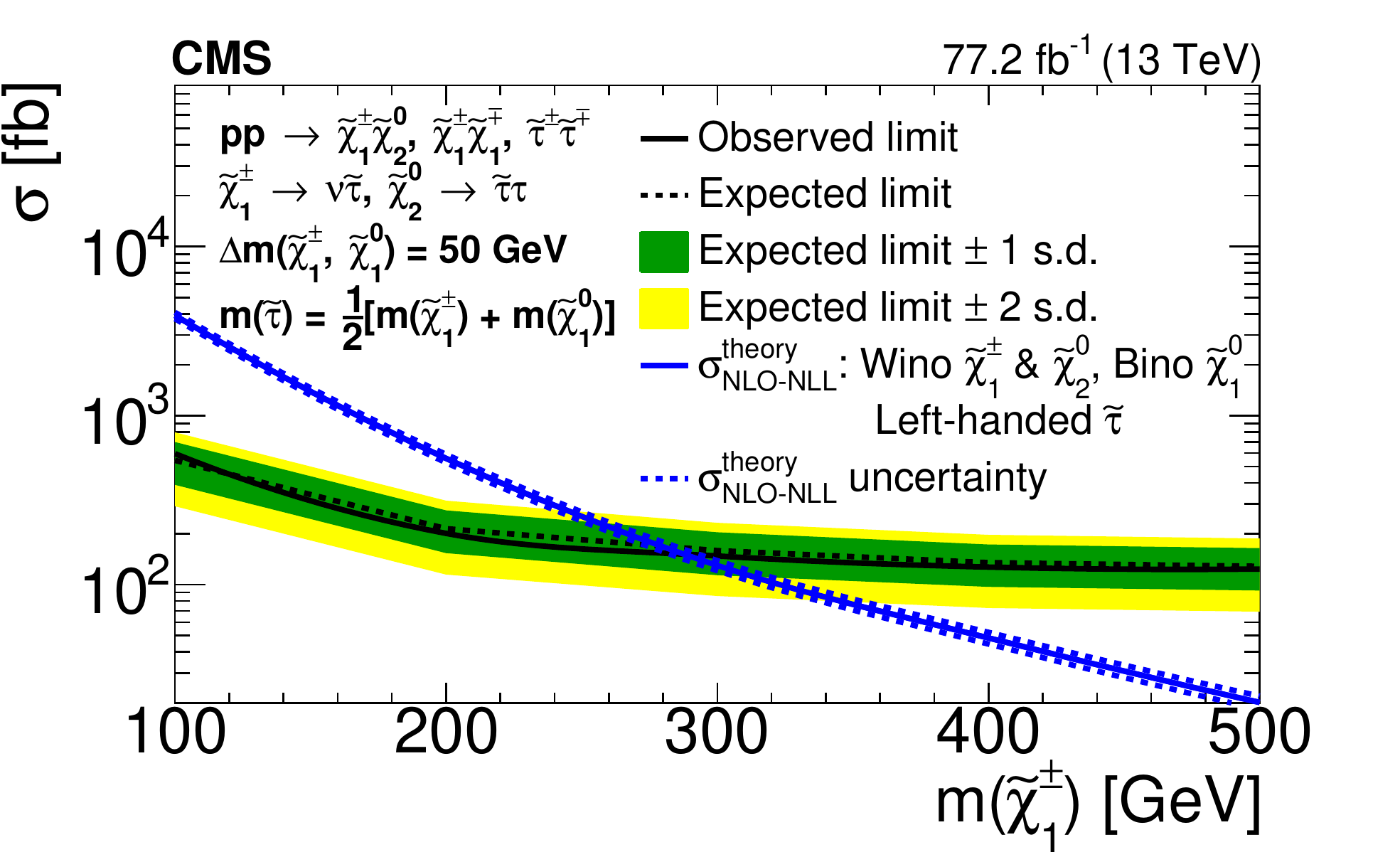}
  \includegraphics[width=0.48\textwidth]{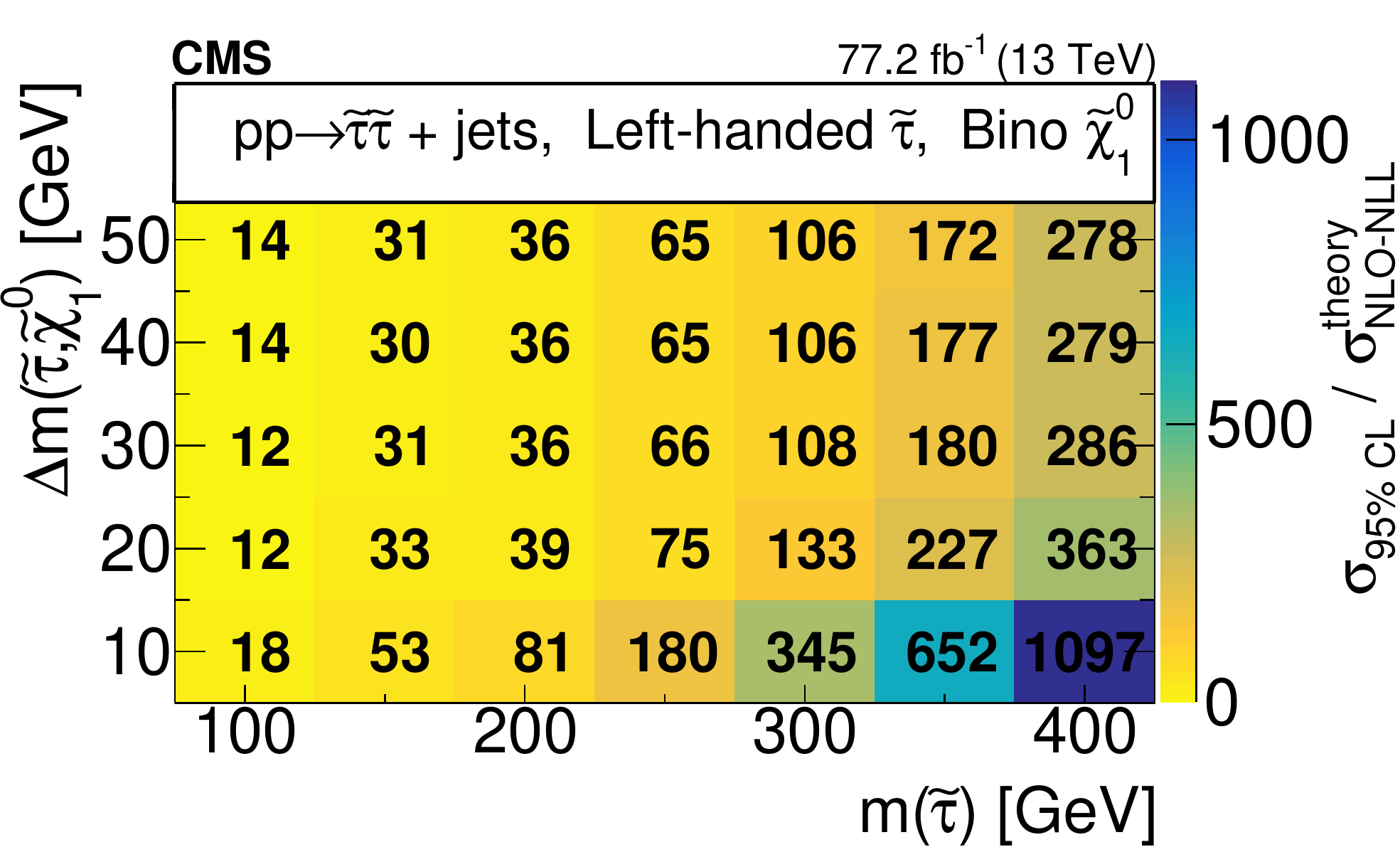}
 \caption{(\cmsLeft)  The 95\% confidence level (\CL) upper limits on the SSM1 production cross sections ($\sigma_{95\%\,\CL}$) as a function of $m(\PSGcpmDo)$. The solid blue
line shows the theoretical cross section, and the dashed blue line its uncertainty.  The observed limit is shown with the solid black line,
while
the expected limit is shown with the dashed black line. The green (yellow) band corresponds to the one (two) standard deviation range about the central value of the expected limit. (\cmsRight)
The ratio of the 95\% \CL upper limit on the direct \PSgt pair production signal cross section in SSM2 to the theoretical cross section, as a function of $m(\PSgt)$ and
$\deltm(\PSgt, \PSGczDo)$.}
 \label{fig:limits}
\end{figure}

No significant excess above the background prediction is observed. The 95\% confidence level (\CL) upper limits are set on the SSM1 signal production cross sections as a function of $m(\PSGcpmDo)$ for fixed $\deltm(\PSGcpmDo,\PSGczDo) = 50\GeV$ and $m(\PSgt) = 0.5 m(\PSGcpmDo) + 0.5 m(\PSGczDo)$ (Fig.~\ref{fig:limits} \cmsLeft). This benchmark is motivated by: $(i)$ LHC searches to date have no sensitivity in these SSM1 compressed spectrum scenarios; and ($ii$) SSM1 scenarios with $\deltm(\PSgt,\PSGczDo) = 25\GeV$ provide the right CA cross section to give a DM relic density consistent with experiment~\cite{Arnowitt:2006bb,Khotilovich:2005gb, Carena:2012gp, Aboubrahim:2017aen, Duan:2018cgb, PhysRevLett.100.231802, VBFPheno, VBFPheno2}. Figure~\ref{fig:limits} \cmsRight shows the observed 95\% \CL  upper limits on the SSM2 signal production cross sections as a function of $m(\PSgt)$ and $\deltm(\PSgt, \PSGczDo)$. The limits are estimated following the modified frequentist construction \CLs method~\cite{Cowan:2010js,Junk,CLs}. Maximum likelihood fits are performed using the final \mT distributions for 2016 and 2017 data to construct a combined profile likelihood ratio test statistic~\cite{Cowan:2010js} in bins of \mT. Systematic uncertainties are represented by nuisance parameters, assuming log-normal priors for normalization parameters, and Gaussian priors for shape uncertainties. Statistical uncertainties in the shape templates are accounted for by the technique described in Ref.~\cite{Barlow:1993dm}. Correlations among the signal and backgrounds have been considered. For example, the uncertainty in the integrated luminosity is treated as fully correlated across signal and backgrounds, while uncertainties from event acceptance variation with different sets of PDFs or variations in the ISR correction factors, in a given \mT bin, are treated as uncorrelated. Uncertainties from the closure tests are treated as uncorrelated. We note that the statistical uncertainty dominates the sensitivity.

For SSM1, we exclude $\PSGczDt/\PSGcpmDo$ with masses below 290\GeV for $\deltm(\PSGcpmDo,\PSGczDo) = 50\GeV$ and $\deltm(\PSGcpmDo, \PSgt) = 25\GeV$. Prior experimental constraints on the
SUSY parameters with these $\deltm(\PSGcpmDo,\PSGczDo)$ and $\deltm(\PSGcpmDo,\PSgt)$ values using non-ISR searches~\cite{CMSEWK, CMSsoftdilepton, Aad:2014nua, ATLASTsoftdilepton,
Khachatryan:2015kxa, SUS17007} have not
exceeded those of the LEP experiments for indirect \PSgt production~\cite{Heister:2001nk, Abdallah:2003xe, Achard:2003ge, Abbiendi:2003ji}. Thus the search presented in this Letter provides the
first results from the LHC to surpass the LEP bound of 103.5\GeV for $m(\PSGcpmDo)$ for such compressed scenarios. For SSM2, small $\PSgt\PSgt$ production cross sections and low signal acceptances
make these scenarios challenging, especially when $\deltm(\PSgt,\PSGczDo) \leq
50\GeV$. For a \PSgt mass of 100 GeV and $\deltm(\PSgt,\PSGczDo) = 30\GeV$, for example, the observed limit is 12 times the theoretical cross section.
It is again noted that the SSM2 results are included in this Letter to highlight
the improved sensitivity in this analysis compared to previous non-ISR searches. A direct comparison with the most sensitive non-ISR search, Ref.~\cite{ATLASTDirectStau}, shows
$\approx\times$4 improvement in the cross section upper limit for the SSM2 scenario with $m(\PSgt) = 150$\GeV and $\deltm(\PSgt,\PSGczDo) = 50\GeV$.

In summary, we have presented a search for compressed supersymmetry. It is the first collider search with exactly one soft, hadronically-decaying tau lepton and missing transverse momentum recoiling against an initial-state radiation jet with high transverse momentum.  The search utilizes data corresponding to an integrated luminosity of 77.2\fbinv collected with the CMS detector in proton-proton collisions at $\sqrts = 13\TeV$. This search targets scenarios where the mass difference ($\deltm$) between the stau (\PSgt) particle and the lightest neutralino (\PSGczDo) is $\leq$50\GeV. This is motivated by models considering \PSgt-\PSGczDo CA to maintain consistency in the relic DM density between particle physics and cosmology. In the context of the minimal supersymmetric standard model, the search considers electroweak production of \PSgt via decays of the lightest chargino (\PSGcpmDo) and the next-to-lightest neutralino (\PSGczDt), and direct production of \PSgt. The data do not reveal evidence for new physics. For a mass splitting $\deltm(\PSGcpmDo,\PSGczDo) = 50\GeV$ and a branching fraction of 100\% for $\PSGcpmDo\to \PSgt\Pgngt \to \tau\PSGczDo\Pgngt$, \PSGcpmDo masses up to 290\GeV are excluded at 95\% confidence level. This sensitivity exceeds that of all other \PSgt searches to date in these scenarios.  The search presented in this Letter provides the first results from the LHC to surpass the LEP bounds.

\begin{acknowledgments}
We congratulate our colleagues in the CERN accelerator departments for the excellent performance of the LHC and thank the technical and administrative staffs at CERN and at other CMS institutes for their contributions to the success of the CMS effort. In addition, we gratefully acknowledge the computing centers and personnel of the Worldwide LHC Computing Grid for delivering so effectively the computing infrastructure essential to our analyses. Finally, we acknowledge the enduring support for the construction and operation of the LHC and the CMS detector provided by the following funding agencies: BMBWF and FWF (Austria); FNRS and FWO (Belgium); CNPq, CAPES, FAPERJ, FAPERGS, and FAPESP (Brazil); MES (Bulgaria); CERN; CAS, MoST, and NSFC (China); COLCIENCIAS (Colombia); MSES and CSF (Croatia); RPF (Cyprus); SENESCYT (Ecuador); MoER, ERC IUT, PUT and ERDF (Estonia); Academy of Finland, MEC, and HIP (Finland); CEA and CNRS/IN2P3 (France); BMBF, DFG, and HGF (Germany); GSRT (Greece); NKFIA (Hungary); DAE and DST (India); IPM (Iran); SFI (Ireland); INFN (Italy); MSIP and NRF (Republic of Korea); MES (Latvia); LAS (Lithuania); MOE and UM (Malaysia); BUAP, CINVESTAV, CONACYT, LNS, SEP, and UASLP-FAI (Mexico); MOS (Montenegro); MBIE (New Zealand); PAEC (Pakistan); MSHE and NSC (Poland); FCT (Portugal); JINR (Dubna); MON, RosAtom, RAS, RFBR, and NRC KI (Russia); MESTD (Serbia); SEIDI, CPAN, PCTI, and FEDER (Spain); MOSTR (Sri Lanka); Swiss Funding Agencies (Switzerland); MST (Taipei); ThEPCenter, IPST, STAR, and NSTDA (Thailand); TUBITAK and TAEK (Turkey); NASU (Ukraine); STFC (United Kingdom); DOE and NSF (USA). \end{acknowledgments}

\bibliography{auto_generated}

\cleardoublepage \appendix\section{The CMS Collaboration \label{app:collab}}\begin{sloppypar}\hyphenpenalty=5000\widowpenalty=500\clubpenalty=5000\vskip\cmsinstskip
\textbf{Yerevan Physics Institute, Yerevan, Armenia}\\*[0pt]
A.M.~Sirunyan$^{\textrm{\dag}}$, A.~Tumasyan
\vskip\cmsinstskip
\textbf{Institut f\"{u}r Hochenergiephysik, Wien, Austria}\\*[0pt]
W.~Adam, F.~Ambrogi, T.~Bergauer, J.~Brandstetter, M.~Dragicevic, J.~Er\"{o}, A.~Escalante~Del~Valle, M.~Flechl, R.~Fr\"{u}hwirth\cmsAuthorMark{1}, M.~Jeitler\cmsAuthorMark{1}, N.~Krammer, I.~Kr\"{a}tschmer, D.~Liko, T.~Madlener, I.~Mikulec, N.~Rad, J.~Schieck\cmsAuthorMark{1}, R.~Sch\"{o}fbeck, M.~Spanring, D.~Spitzbart, W.~Waltenberger, C.-E.~Wulz\cmsAuthorMark{1}, M.~Zarucki
% ...
\vskip\cmsinstskip
\dag: Deceased\\
1:  Also at Vienna University of Technology, Vienna, Austria\\
% ...
\end{sloppypar}
\end{document}